\documentclass{PoS}

\title{Matrix elements of heavy-light mesons from a fine lattice}

\ShortTitle{Heavy-light mesons from a fine lattice}

\author{\speaker{A.~Ali~Khan},A.~Al-Haydari$^a$,
V.~M.~Braun$^b$, S.~Collins$^b$,
M.~G\"ockeler$^b$, G.~N.~Lacagnina$^c$, M.~Panero$^{b, d}$,
A.~Sch\"afer$^b$, and G.~Schierholz$^{b, e}$\\

\noindent $^a$ 
 Department of Physics, Faculty of Science, Taiz University, Taiz, Yemen Republic\\
$^b$ 
 Institute for Theoretical Physics, University of
Regensburg, 93040 Regensburg, Germany\\
$^c$ 
 INFN, Sezione di Milano, 20133 Milano, Italy\\
$^d$ 
 Institute for Theoretical Physics, ETH Z\"urich, 8093 Z\"urich, Switzerland\\
$^e$ 
 Deutsches Elektronen-Synchrotron DESY, 22603 Hamburg,
  Germany
}

\abstract{We performed a calculation of matrix elements of
heavy mesons on a quenched lattice,  generated with Wilson gauge 
fields at $\beta = 6.6$ with  a lattice size of $40^3\times  80$ and a 
lattice spacing $a^{-1} \simeq 5$ GeV determined from the Sommer parameter 
$r_0 = 0.5$ fm. We use a 
non-perturbatively $O(a)$ improved Wilson fermion action and improved currents.

We have calculated the charmonium spectrum as well as form factors of 
semileptonic decays of 
pseudoscalar heavy-light mesons containing a $c$ or a $b$ quark to 
pseudoscalar light mesons through a vector current:
\[
\langle P (p) \vert V^\mu \vert H(p_H)\rangle =  \frac{m_H^2 - m_P^2}{q^2} q^\mu 
f_0 (q^2) + \left( p^\mu_H + p^\mu - \frac{m_H^2 - m_P^2}{q^2}q^\mu  \right) f_+(q^2)\,,
\]
where $p$ and $m_P$ are the momentum and the mass of the light meson respectively, and $p_H$ and
$m_H$ the momentum and the mass of the heavy meson respectively. $q = p_H - p$ is the momentum
transfer. $V^\mu$ denotes a local vector current.
A comparison with other lattice calculations for the decay $B \rightarrow \pi 
l \nu$ is shown in Figure \protect\ref{fig:decays}.

For the decay constant of the $J/\psi$ meson we find a preliminary value of
$f_V = 341(10)$ MeV, using the definition
\[
\langle 0 | V_j | J/\psi \rangle = f_V m_{J\psi} \epsilon_j(\lambda),
\]
where $\epsilon_j(\lambda)$ is the polarization vector of the $J/\psi$.
We obtain the following values for mass splittings of charmonium states:
$\Delta M(J/\psi-\eta_c) = 74(2)$ MeV and $\Delta M(\chi_{c1}-J/\psi) = 394(15)$
MeV, where the errors are only statistical.}

\FullConference{8th Conference Quark Confinement and the Hadron Spectrum \\
		 September 1-6 2008\\
		 Mainz, Germany}

\begin{document}

The details of our calculation of the matrix elements and the results for the
form factors at $q^2 = 0$ are given in Ref.~\cite{alhaydari}. 
\begin{figure}
\centerline{
\includegraphics[width=.60\textwidth]{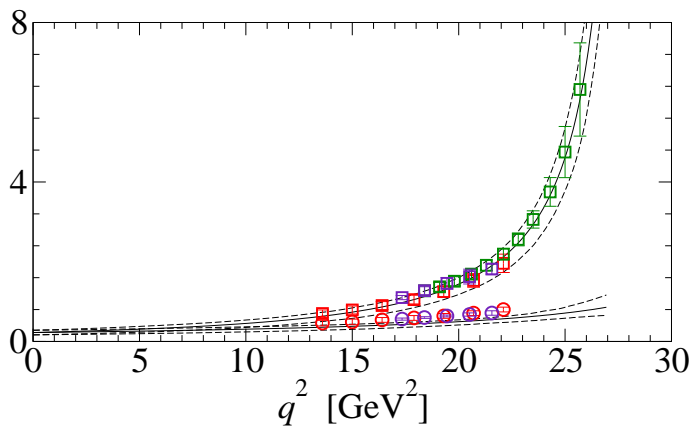}
\hspace{.04\textwidth}
}
  \caption{The solid lines denote our results for the form factors ($f_+$: 
upper line, $f_0$: lower line) of the decay $B \rightarrow \pi l \nu $. The 
dashed lines denote our error bounds.  
The  squares denote $f_+$, circles denote $f_0$ from other recent lattice 
calculations (red: quenched, \protect\cite{Abada:2000ty},
green: $N_f = 2+1$, \protect\cite{Bailey:2008wp}, magenta:  $N_f = 2+1$,
\protect\cite{Dalgic:2006dt})}. 

  \label{fig:decays}
\end{figure}
We observe a relatively good agreement of our form factors with other lattice  
calculations.

Our results for the charmonium spectrum are in very good agreement with
a previous quenched calculation at the same lattice spacing \cite{Choe:2003wx}
and in agreement with a recent  calculation with two heavy flavors
\cite{Ehmann:2007hj}. Since we work on fine lattices where discretization
effects are under good control for charmonia it is also of interest to 
calculate the charmonium decay constants.
Our result for the decay constant
of the $J/\psi$ is lower than the result of 399(4) MeV from quenched 
anisotropic lattices of \cite{Dudek:2006ej} and of 413(40) MeV from lattices 
with $N_f = 2$ of \cite{Dimopoulos:2008ee}. The experimental value is 411(7) 
MeV.

\noindent {\bf Acknowledgement:} We thank J.~Simone for useful discussions.

\end{document}